# *Fortress and Gatekeeper*: Theorizing Transitive Trust in Third-Party Cybersecurity Risk Governance

**Yijun Chen, Macquarie University, NSW, Australia**
**Misita Anwar, Monash University, Swinburne University of Technology, VIC, Australia**

## Abstract

Third-party vendors, including analytics platforms, cloud services, identity providers, and software suppliers, have become embedded in digital service delivery. These arrangements enable scale and specialization, but they also move customer data and security-relevant practices into environments that customers rarely see, select, or evaluate. This paper examines this problem through a document analysis of the November 2025 OpenAI–Mixpanel security incident. The incident is used as an illustrative case to examine how a security event in a vendor environment can become a governance and accountability problem for the focal organisation that holds the customer relationship. Drawing on organisational trust research and agency theory, the paper argues that third-party cybersecurity risk involves both a trust relationship and a delegation problem. Customers place trust in the visible service provider, while the provider depends on vendors whose security practices are only partially visible and controllable. To explain this relationship, the paper develops the concept of transitive trust, in which customer trust in a digital service depends on the security practices of third-party vendors authorized by that service provider. The paper then presents the Fortress and Gatekeeper framework, which explains why cybersecurity governance boundaries should be understood in terms of trust and data flows, rather than formal organizational ownership alone. The analysis develops four propositions concerning vendor integration, metadata exposure, vendor assurance, and data proliferation. The paper contributes to cybersecurity governance scholarship by explaining how delegated data processing can create customer-facing accountability and by identifying implications for vendor tiering, data classification, contractual design, continuous assurance, and data minimization.

## 1. Introduction

Digital service organizations increasingly depend on external providers for cloud infrastructure, analytics, identity management, customer support, and security monitoring (Benaroch & Fink, 2021). These arrangements accelerate product delivery and enable specialization, but they also distribute customer data and operational responsibility across organizational boundaries that customers cannot select or observe (Boyson et al., 2022). When vendors experience a security incident, the consequences propagate to the focal organization whose users recognize it as the trusted counterparty. In response, cybersecurity governance therefore no longer terminates at the perimeter of the organization that holds the customer relationship; it extends into an ecosystem of suppliers whose controls shape downstream exposure. For example, the November 2025 OpenAI–Mixpanel incident, in which unauthorized access occurred within a vendor environment rather than the focal platform, illustrates how this displacement of technical risk and customer-facing accountability has become a structural feature of contemporary digital service delivery (Kovacs, 2025).

The shift from perimeter protection to ecosystem governance is reflected in leading frameworks. The National Institute of Standards and Technology (NIST) Cybersecurity Framework (CSF) 2.0 situates cybersecurity supply chain risk management within the governance function and links supplier oversight to enterprise risk management, policy, and accountability (National Institute of Standards and Technology, 2024). NIST Cybersecurity Supply Chain Risk Management (C-SCRM) guidance further defines supply chain risk as a life-cycle problem across interconnected information and communication technology (ICT) supply chains (National Institute of Standards and Technology, 2024). Meanwhile, the International Organization for Standardization/International Electrotechnical Commission (ISO/IEC) 27001 similarly frames supplier relationships, risk assessment, and continual improvement as integral to (ISO-International Organization for Standardization / IEC-International Electrotechnical Commission, 2022). Industry evidence aligns with these normative shifts: the World Economic Forum (2025) reports that a majority of large organizations now identify supply chain challenges as the leading barrier to cyber resilience.

Complementary scholarship provides theoretical grounding. Research on digital service ecosystems emphasizes that firms create value through networks of specialized providers rather than through vertically integrated operations (Hofmann & Osterwalder, 2017). Existing research demonstrates that attackers exploit dependencies and suppliers to bypass the defences of more mature targets (Ohm et al., 2020). In this sense, agency theory clarifies why delegation produces information asymmetry between focal organizations and their vendors (Eisenhardt, 1989), while organizational trust research distinguishes direct trust from system-level trust in intermediaries (Mayer et al., 1995). Privacy scholarship further cautions that data sensitivity is context-dependent rather than fixed by category alone (Coombs et al., 2020). Together, these literatures provide partial accounts of vendor-mediated risk but do not converge on a unified mechanism that links customer trust, delegated processing, and accountability.

Despite these advances, the conceptual vocabulary available for analysing third-party cyber risk remains underdeveloped (Liu & Babar, 2026). Vendor questionnaires, certifications, contractual clauses, and annual audits are useful governance tools, but they do not explain the relational structure of trust that arises when a customer interacts with a visible organization that relies on largely invisible vendors (Boyson et al., 2022). Customers rarely select or even identify the analytics providers, support processors, identity services, or infrastructure vendors behind a digital platform, yet their data and security exposure are shaped by these choices. Existing frameworks treat vendor assurance primarily as a procedural or compliance activity and offer limited theoretical articulation of the mechanism through which a vendor incident becomes a first-party accountability event (Creazza et al., 2021). The problem is further compounded by data proliferation across analytics, experimentation, and support platforms, each of which adds retention, export, and notification obligations that must be reconciled during incidents. In particular, with AI systems increasingly trained on, fine-tuned with, or served by vendor-managed data pipelines, the volume and granularity of data flowing outside the focal organization's direct custody have expanded substantially (Luna et al., 2026). Without an explicit construct for this delegated trust structure, governance practice risks drawing boundaries along organizational ownership rather than along the trust and data flows that actually determine downstream exposure.

Against this backdrop, the present study asks *how organizations should govern cyber risk when customer trust and customer data move through third-party systems*. Using document analysis (Bowen, 2009) of the OpenAI–Mixpanel incident, together with governance standards and relevant scholarship, the study introduces the construct of transitive trust and develops our theoretical framework, which we call the Fortress and Gatekeeper framework, to reconceptualize cyber governance boundaries as tracing trust and data flows rather than organizational ownership alone. Drawing on the agency theory proposed by Eisenhardt (1989), this study maps the governance surface of third-party cyber risk and articulates implications for vendor tiering, data minimization, contractual design, and continuous assurance.

The paper makes three contributions. First, it introduces transitive trust as a construct to explain how customer trust in a focal digital service depends on vendors that customers rarely see, select, or evaluate. Second, we adopt agency theory to explain why these vendor relationships are difficult to govern: focal organisations remain accountable to customers, but they can only partially observe and control the security practices of the vendors that handle customer data or support service delivery. Third, we develop the Fortress and Gatekeeper framework to show how customer trust, delegated data processing, vendor opacity, assurance over time, and first-party accountability interact in third-party cybersecurity risk governance.

The remainder of this paper presents the theoretical background, describes the document-analysis method, analyses the case, develops the propositions, and discusses contributions, boundary conditions, and directions for future research.

## 2. Theoretical Background

### 2.1 From perimeter to ecosystem: Third-party cyber risk

Third-party cyber risk refers to the possibility that an organization's confidentiality, integrity, availability, privacy obligations, operations, or reputation will be harmed by the practices or incidents of an external provider (The Financial Stability Board, 2023). The risk is relational: it emerges when data, access privileges, processing activities, or operational dependencies are distributed across organizations (Boyson et al., 2022). For example, a vendor that receives identifiable user data or supports a critical workflow becomes part of the focal organization's practical security boundary, even though it remains legally and operationally separate (Kunnathur, 2015; Liu & Babar, 2026; Shukla et al., 2022). Research on digital service ecosystems emphasizes that contemporary firms create value through networks of specialized providers rather than through vertically integrated operations alone (Hofmann & Osterwalder, 2017). Cybersecurity research reaches a parallel challenge: attackers can exploit dependencies, software components, and suppliers to bypass the defences of more mature targets (Ohm et al., 2020). The governance challenge is, therefore, whether organisations can effectively identify, prioritize, monitor, and mitigate the risks introduced by the systems upon which it relies (Keskin et al., 2021). In other words, a firm's exposure no longer ends at its own perimeter; its data and operations extend into systems that it neither owns nor controls. Consequently, the boundary that is crucial for governance is defined by trust and data flow, rather than by legal ownership (Santos & Eisenhardt, 2005).

### 2.2 Delegation and information asymmetry

According to Eisenhardt (1989), a foundational assumption of agency theory is that information is distributed asymmetrically between the parties to a contractual relationship, with the agent holding private knowledge that the principal can access only imperfectly and at cost. When a focal organization engages a vendor to perform a business or technical function, the arrangement is not simply an exchange of service for payment; it is a delegation of action to an agent whose conduct the principal cannot fully observe (Jensen & Meckling, 1976). Agency scholarship distinguishes two forms of the resulting information asymmetry — *hidden information*, concerning attributes of the agent that are known at contracting but difficult for the principal to verify, and *hidden action*, concerning conduct that unfolds after contracting but is similarly unobservable (Eisenhardt, 1989). Cybersecurity governance is acutely exposed to both. In particular, a vendor's control maturity, incident history, staffing practices, and patching discipline are largely private at the point of engagement, while its ongoing investment in defensive controls is shaped by cost–benefit calculations that unfold outside the client's line of sight (Creazza et al., 2021). The configuration is a textbook moral hazard. The vendor holds private incentives to minimize friction, overhead, and expenditure, while the distributional consequences of under-investment — reputational damage, customer attrition, and regulatory scrutiny — fall disproportionately on the focal organization whose brand mediates the customer relationship (Menon & Siponen, 2020). The costs and benefits of vendor effort are thus borne by different parties, and the party bearing the cost is not the party best positioned to monitor the effort.

In addition, contractual terms, third-party certifications, and periodic audits are the standard instruments for narrowing this gap. Within agency theory, these devices are properly understood as monitoring and bonding mechanisms (Jensen & Meckling, 1976): they constrain the agent's discretion, generate verifiable signals of compliance, and establish contractually enforceable consequences for documented failure. Their effectiveness, however, is contingent on the ongoing production of credible evidence, a condition rarely satisfied by point-in-time attestation. Certifications describe a control environment at a single point in time, contractual audit rights are exercised infrequently, and questionnaire responses are self-reported by the very party whose conduct is under examination (Lins et al., 2018). The result is a governance regime that documents the existence of controls without sustaining the continuous verification that agency theory's monitoring logic requires.

### 2.3 Trust, transitive trust, and the Fortress–Gatekeeper structure

Trust research distinguishes direct trust in an actor from trust placed in the systems or intermediaries that support that actor (Mayer et al., 1995). In digital service ecosystems, customers direct their trust toward the visible platform, brand, or service provider. That trust is then extended, through the focal organization's vendor choices, to processors the customer neither selected nor observed. The customer does not approve of these choices, but the customer's data and security exposure are shaped by them. In this sense, we term this structure transitive trust: trust in a focal organization is implicitly delegated to the vendors that the organization selects and governs. According to prior literature, the construct of transitive trust synthesizes insights from three distinct theoretical streams. Grounded in trust research, previous work adopts the distinction between trustee-specific trust and institution-based trust (McKnight et al., 2002), building on the premise that trust is embedded in conditions of risk, interdependence, and ongoing relationships (Li et al., 2024; Rousseau et al., 1998).

To account for the structural vulnerabilities arising from such relationships, the construct draws on agency theory to frame the problem of information asymmetry in nested delegation. Specifically, customers, as ultimate principals, place trust in a focal organization that may subsequently delegate data processing, analytics, infrastructure provision, access management, or security monitoring to third-party vendors. This creates a chain of delegated relationships in which visibility and control may diminish as data and operational responsibilities move from the focal organisation to third-party vendors and, in some cases, to vendor employees (Dattathrani & De', 2023; Eisenhardt, 1989). Finally, cyber supply chain research highlights the practical consequence of this architecture: cybersecurity risks often emerge through interdependencies among organizations, systems, and third-party providers rather than through isolated organizational boundaries (Sadeghi R. et al., 2024). By integrating these perspectives, transitive trust formalizes a mechanism implicit across these literatures yet under-theorized in cybersecurity governance for third-party digital service dependencies. As such, we formalize this structure through the *Fortress and Gatekeeper framework*. The fortress represents the focal organization and its visible internal defences. The gatekeepers represent vendors that process data, operate analytics, provide infrastructure, manage access, or support user-facing functions. A fortress can be technically robust while remaining exposed through a gatekeeper that receives data or holds operational privileges. The framework's core theoretical proposition is that cyber governance boundaries follow trust and data flows, not organizational ownership alone. To further explain this in detail, **Figure 1** summarizes this logic.

**Figure 1.** The Fortress and Gatekeeper framework: transitive trust, delegated data processing, and governance controls.

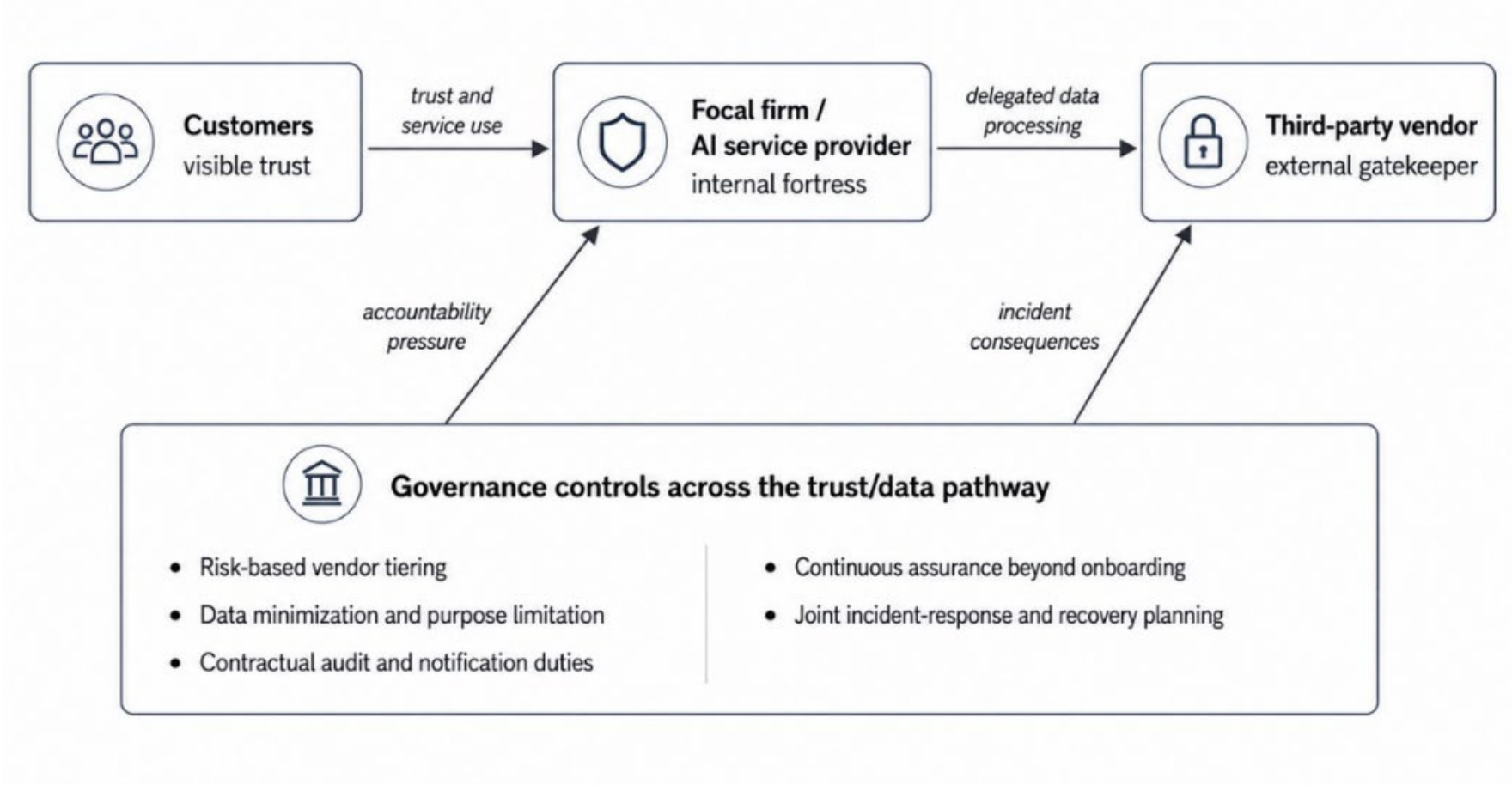

## 3. Research Method

### 3.1 Document analysis as methodological approach

This study uses document analysis as its research method. Document analysis is a qualitative approach in which texts are systematically reviewed and interpreted to develop understanding, surface themes, and examine organizational meaning (Bowen, 2009); Morgan, 2022). This approach is appropriate because the study does not seek to conduct a technical forensic investigation of the OpenAI–Mixpanel incident. Rather, it uses publicly available incident documents, governance standards, and relevant scholarship to interpret how a vendor-side incident can reveal broader governance mechanisms and accountability relationships. Thus the purpose is to use a publicly documented case to refine theoretical concepts that may help explain similar third-party cybersecurity governance problems (Morgan, 2022).

### 3.2 Document corpus and selection criteria

Documents were selected against three criteria: relevance to the OpenAI–Mixpanel incident, relevance to third-party cyber risk governance, and public accessibility. The corpus comprised four categories, summarized in Table 1. Incident-specific documents including OpenAI's public notice and Mixpanel's public response, provided the factual basis for the case timeline, data categories, and organizational responses. Governance frameworks, including NIST CSF 2.0, NIST C-SCRM guidance, and ISO/IEC 27001, provided normative criteria for interpreting third-party risk governance. Industry context was supplied by the World Economic Forum's Global Cybersecurity Outlook 2025, which situates supply chain vulnerabilities within broader cyber-resilience concerns. Academic literature on document analysis, agency theory, organizational trust, privacy, and supply chain attacks grounded the conceptual interpretation. Official organizational statements were prioritized for case facts because they provide direct accounts of the disclosed incident; standards and frameworks were prioritized for governance interpretation because they define expected practices for enterprise and supplier risk management; academic sources situated the analysis within established theory.

**Table 1.** Document corpus used in the analysis

| Document category | Examples | Analytical purpose |
|---|---|---|
| Incident-specific documents | OpenAI incident notice; Mixpanel public response | Establish timeline, affected data categories, exclusions, and disclosed organizational responses |
| Governance frameworks | NIST CSF 2.0; NIST C-SCRM guidance; ISO/IEC 27001 | Interpret vendor oversight, supply chain risk, policy, roles, monitoring, and response expectations |
| Industry context | World Economic Forum Global Cybersecurity Outlook 2025 | Situate the case within broader cyber-resilience and supply chain risk trends |

| Academic literature | Document analysis; agency theory; organizational trust; privacy; supply chain attack literature | Support method, theory development, and interpretation of governance mechanisms |
|---|---|---|

### 3.3 Analytical procedure

The analysis proceeded through three iterative stages. In the first stage, factual statements were extracted from incident documents, including the system in which the incident occurred, the data categories involved, excluded data categories, timeline information, and response actions. In the second stage, deductive codes were developed from principal-agent theory, organizational trust, and cyber supply chain risk management. These codes included delegated processing, information asymmetry, customer-facing accountability, vendor monitoring, metadata exposure, and data minimization. In the third stage, inductive coding identified recurrent governance themes not anticipated by the initial framework, following a thematic logic consistent with established qualitative coding approaches (Braun & Clarke, 2006). The iterative movement between theoretically informed and emergent interpretation produced four analytical themes—expanded attack surface, actionable metadata, point-in-time assurance limits, and data proliferation—which we subsequently elevate to theoretical propositions (Section 5).

### 3.4 Trustworthiness and boundary considerations

To strengthen credibility, claims were triangulated across document types. Incident facts from OpenAI were compared with Mixpanel's response, and both were interpreted against NIST and ISO governance expectations. Transferability is supported by an explicit description of the case context; dependability is supported by a documented coding procedure; confirmability is supported by the use of public documents that can be independently examined by other researchers. The analysis remains bounded by the available public record. It does not infer undisclosed contractual terms, internal security assessments, root-cause details, or organizational decision processes. The study is appropriately read as theory-refining rather than theory-testing: it uses a single, well-documented case to articulate and refine a conceptual framework that future empirical work can test at scale.

## 4. Case Context: The OpenAI–Mixpanel Incident

The OpenAI–Mixpanel incident became public in late November 2025. Mixpanel stated that on November 8, 2025, it detected a smishing campaign and activated incident-response processes, including securing impacted accounts, revoking active sessions, rotating compromised credentials, blocking malicious IP addresses, registering indicators of compromise, performing global employee password resets, engaging a third-party forensics firm, reviewing authentication and export logs, and engaging law enforcement and external cybersecurity advisors (Kovacs, 2025). According to OpenAI, on November 9, 2025, Mixpanel became aware that an attacker had gained unauthorized access to part of its systems and exported a dataset containing limited customer-identifiable and analytics information; Mixpanel shared the affected dataset with OpenAI on November 25, 2025 (OpenAI, 2025).

In addition, OpenAI emphasized that the incident was limited to Mixpanel's systems and did not involve unauthorized access to OpenAI infrastructure. Affected information may have included names associated with accounts, email addresses, approximate coarse location based on browser data, operating system and browser information, referring websites, and organization or user identifiers. OpenAI stated that chat content, prompts, responses, API requests, API usage data, passwords, credentials, API keys, payment details, government IDs, session tokens, authentication tokens, and other sensitive parameters for OpenAI services were not affected (OpenAI, 2025). OpenAI's disclosed response included removing Mixpanel from production services, reviewing affected datasets, notifying impacted organizations and users, monitoring for misuse, terminating its use of Mixpanel, conducting expanded security reviews across the vendor ecosystem, and elevating security requirements for partners and vendors (OpenAI, 2025). These actions are analytically significant: a third-party incident triggered substantive first-party governance obligations even though the focal organization's own infrastructure was not breached.

The case therefore exemplifies what cyber risk in a decoupled world (Baldoni, 2022) where technical compromise, data stewardship, and customer accountability occur in different organizational locations. The vendor environment was the technical site of unauthorized access; OpenAI remained the organization users recognized as responsible for communication, mitigation, and future assurance. This decoupling is the empirical substrate on which the Fortress and Gatekeeper framework is built.

**Table 2.** Public timeline of the OpenAI–Mixpanel incident

| Date | Disclosed event | Governance relevance |
|---|---|---|
| November 8, 2025 | Mixpanel detected a smishing campaign and initiated incident response. | Supplier-side identity and social-engineering risks propagate to downstream clients. |
| November 9, 2025 | Mixpanel became aware of unauthorized access and dataset export. | Vendor system compromise is separated from first-party customer accountability. |
| November 25, 2025 | Mixpanel shared the affected dataset with OpenAI. | Notification speed, evidence sharing, and customer-impact assessment become governance constraints. |
| November 26–27, 2025 | OpenAI and Mixpanel issued public communications. | Incident communication and trust repair emerge as governance functions. |
| After disclosure | OpenAI removed Mixpanel from production and expanded vendor security reviews | Post-incident remediation extends to ecosystem-level assurance. |

## 5. Case analysis and proposition development

This section uses the OpenAI–Mixpanel incident as an illustrative case to refine theory on third-party cybersecurity governance. Our analysis reveals that the OpenAI–Mixpanel incident operates through a *nested agency structure* in which end users delegate trust to OpenAI; OpenAI delegates processing to Mixpanel; and Mixpanel, in turn, delegates operational duties to its employees. The public record shows that the technical incident occurred in Mixpanel's systems rather than OpenAI's infrastructure. The case illustrates how each link is a principal–agent relationship whose conduct the upstream principal cannot continuously observe. The disclosed facts of the case are consistent with this interpretation. According to Mixpanel's chief executive Jen Taylor, the company "detected a smishing campaign" against its employees on 8 November 2025 and activated incident-response processes (Taylor, 2025). On the other hand, OpenAI's public notice, published on 26 November 2025, states that *the intrusion was not a breach of OpenAI's systems* (OpenAI, 2025) and reports that on 9 November 2025 Mixpanel had become aware of unauthorised access to part of its environment, with the attacker exporting a dataset of customer-identifiable and analytics information. The public reports describe a coordinated two-party disclosure and characterise the incident as arising from a smishing campaign that affected Mixpanel's employee credentials. Analytically, this is important because the incident occurred within the vendor environment, while many of the consequences were borne by the focal organisation that maintained the customer relationship. OpenAI was responsible for reviewing the affected dataset, communicating with impacted users and organisations, monitoring potential misuse, reassessing the vendor relationship, and managing reputational exposure, even though its own systems were not reported to have been breached. This separation between the location of technical compromise and the location of customer-facing accountability is the central governance issue examined in the analysis that follows. Therefore, the following analysis develops four propositions. Each proposition identifies a governance mechanism made visible by the case and links it to the broader integrated framework of transitive trust.

### 5.1 Proposition 1: Vendor delegation introduces hidden-information risk that expands the effective attack surface

From the perspective of agency theory, vendor integration can be understood as a delegated relationship in which the focal organisation depends on a vendor whose internal security conditions cannot be fully observed (Eisenhardt, 1989). At the point of engagement, our results reveal that the focal organization cannot fully verify the vendor's operational posture—its employee population, social-engineering resilience, privilege-separation discipline, data-retention practices, or the security of its own downstream dependencies. These attributes are *hidden information*: properties of the agent that shape exposure but resist inspection by the principal. The Mixpanel case makes the structural consequence visible. The attack vector was not a software flaw but a smishing campaign against Mixpanel staff —a vulnerability rooted in the human layer of the agent's organisation, which onboarding questionnaires and SOC-type attestations cannot meaningfully surface.

The scale of the hidden-information problem is ecosystem-wide; in particular, Mixpanel serves a large corporate customer base, and the November incident propagated beyond OpenAI to other integrated clients, including CoinTracker (Ilascu, 2025). Industry survey evidence further indicates that every one of 544 websites scanned in the third quarter of 2025 relied on third-party vendors, with 95 per cent running analytics trackers (Tsarynny, 2025). Each such delegation extends the principal's effective attack surface along dependencies that the customer neither selected nor can observe. Crucially, the risk identified in P1 is also structural. When a vendor employee is compromised, the consequences can extend to multiple client organisations that have delegated data processing to that vendor. This occurs because each client depends on security conditions within the vendor environment that it cannot fully observe or control. Therefore, we theorise this structural vulnerability as a **transparency asymmetry** and formalise it as Proposition 1:

> ***P1:*** *The cybersecurity governance boundary of a focal organisation expands as it integrates more vendors that handle customer data or support critical services. Each vendor relationship introduces security conditions that the focal organisation cannot fully observe, meaning that a vendor incident can create first-party accountability even when the focal organisation's own infrastructure is not compromised.*

### 5.2 Proposition 2: Metadata exposure reveals the limits of conventional data-sensitivity categories

Second, OpenAI's public notice distinguished between the information that may have been exposed and the information that was not affected. The exposed data may have included names, email addresses, approximate location, browser and operating-system information, referring websites, and user or organisation identifiers. By contrast, OpenAI stated that chats, prompts, credentials, API keys, payment details, session tokens, and government IDs were not affected (OpenAI, 2025). This distinction reflects a common logic in vendor governance, where data are often classified according to sensitivity levels that guide risk assessment, breach response, notification duties, and contractual protections.

Agency theory suggests that contractual and assurance mechanisms are only as effective as the risks they are able to specify and monitor. Where certain forms of data exposure fall outside established sensitivity categories, the focal organisation may still bear residual loss if those data can be used in harmful ways (Jensen & Meckling, 1976). The Mixpanel incident illustrates this issue. Although exposed information did not include passwords, API keys, payment details, or user-generated content, the affected metadata may still be adversarially actionable. A phishing email that already knows the target's name, employer, rough location, and OpenAI organisation identifier—and that can credibly reference the Mixpanel incident itself as a pretext for "key rotation" or "incident review"—is substantially more persuasive than an untargeted attempt (Prasad, 2025). For example, an attacker could use such information to craft messages that appear relevant to a user's organisation, platform use, or security concerns following the incident. Browser and operating-system fingerprints further support exploit selection, and organisation identifiers enable adversaries to map which firms use the OpenAI API for downstream targeting of those firms' employees (Prasad, 2025). The governance problem is not whether OpenAI's contract with Mixpanel was inadequately drafted. Rather, the case highlights a broader specification gap where conventional data-sensitivity categories may not fully capture the practical ways in which exposed metadata can be combined, contextualised, and weaponised by adversaries.

> ***P2:*** *The governance risk associated with a vendor data exposure is a function not of the sensitivity class named in the contract but of adversarial actionability—the extent to which the exposed data enables targeting, impersonation, reconnaissance, or linkage—so that principals bear residual loss to the degree that contractual data taxonomies diverge from adversary-relevant attributes.*

### 5.3 Proposition 3: Point-in-time assurance weakens as vendor conditions change over time

Agency theory distinguishes hidden information (pre-contractual) from *hidden action* (post-contractual). Hidden action refers to conduct that occurs after delegation and that the principal cannot fully observe, even though it may affect the principal's exposure (Eisenhardt, 1989). Cybersecurity governance is especially vulnerable to this problem because a vendor's ongoing investment in phishing-resistant authentication, employee training, access reviews, patch discipline, and incident preparation is shaped by operational incentives the principal cannot see and that shift continuously over the life of the relationship (Creazza et al., 2021). The standard *monitoring* instruments, such as onboarding questionnaires, annual SOC 2 Type 2 reports, ISO/IEC 27001 certifications, periodic penetration tests, and attestation letters, are structurally mismatched to this form of asymmetry (Kitsios et al., 2023). Each produces a verifiable signal at a single moment, yet the hidden action unfolds continuously between observation points. The Mixpanel incident exemplifies the mismatch. The proximate cause was a smishing campaign that succeeded against employee credentials (Mixpanel, 2025)—precisely the behavioural-layer vulnerability that no annual control attestation reliably surfaces, because its realisation depends on what an individual employee does at a particular moment in response to a particular message.

Our analysis highlights that this empirical timeline deepens the point. For example, Mixpanel detected the smishing campaign on 8 November, identified the exported dataset on 9 November, and shared the affected dataset with OpenAI on 25 November; OpenAI then notified impacted users within 48 hours of receiving the dataset (Kovacs, 2025). The approximate time interval between agent-side detection and principal-side evidence receipt is itself a monitoring gap that agency theory predicts: the principal cannot compress the discovery-to-disclosure interval because evidence production is controlled by the very agent whose conduct is under examination. This is not a failure of contractual diligence but the structural decay of assurance between periodic observation points. The monitoring gap, moreover, does not close at the point of notification. On 19 December 2025, OpenAI amended its original notice to clarify that a limited number of ChatGPT users—not only API users—had been affected, indicating that the "*principal's scope determination continued to evolve for approximately three weeks after public disclosure*". Agency theory predicts this pattern: when the agent controls evidence production, the principal's capacity to resolve scope is bounded by the agent's own investigative pace and by the principal's independent access to relevant logs, both of which are structurally constrained in a delegated processing relationship. Therefore, we theorise this vulnerability as a **monitoring gap** and formalise it as Proposition 3.

*P3: Point-in-time vendor assurance becomes less reliable as vendor conditions change over time. Its governance value weakens when current security exposure depends on actions, behaviours, evidence, or operational conditions that the focal organisation cannot continuously observe or verify.*

### 5.4 Proposition 4: Data proliferation increases the first-party accountability burden

Each additional processor the focal organisation engages is a distinct principal–agent relationship, with its own information asymmetries, bonding instruments, and monitoring overhead. Agency theory frames the total cost of delegation as the sum of *monitoring costs* (borne by the principal), *bonding costs* (borne by the agent), and *residual loss* (borne by the principal) (Jensen & Meckling, 1976). As organisations add analytics, experimentation, support, marketing, identity, and infrastructure vendors, customer data may be copied, logged, retained, exported, or processed across multiple third-party environments. This increases the complexity of understanding and responding to incidents. The OpenAI-Mixpanel incident illustrates this issue under incident conditions. The affected records belonged to OpenAI's users but resided in Mixpanel's environment; OpenAI therefore had to rely on the agent for scope determination, identification of affected records, and evidence production before it could meet its own notification, communication, and remediation obligations (OpenAI, 2025). The case shows how first-party accountability can depend on evidence and data held by a third-party vendor.

Our results show the December 19 scope amendment renders this dependence empirically visible. The distinction between "API users" and the latter clarified "ChatGPT users who submitted help center tickets or were logged into platform.openai.com" maps onto distinct data flows into the same vendor's environment. These flows, including product telemetry, help-centre telemetry, and authenticated browsing of the API platform, constitute three separate channels, each with its own logging configuration and contributing a different subset of records to the exported dataset. Viewed through the lens of agency theory, each channel represents a separate pathway through which customer data enters the vendor's systems. As these pathways multiply, the work required to determine the scope of an incident also increases. This work falls largely on the focal organisation, even when the affected data reside in the vendor environment. Data proliferation thus compounds not only across vendors but also within a single vendor, wherever multiple customer touchpoints feed the same processor. Industry commentary concurs on the direction of travel: as AI platforms push more telemetry, experimentation, and support data into vendor ecosystems, the monitoring load on the principal grows faster than any individual vendor's bonding mechanism can absorb (Tsarynny, 2025). The resulting pattern is **agency-cost concentration**. The party best positioned to observe the agent's conduct is the agent itself; the party bearing most of the reputational, notification, and remediation costs when that conduct fails is the principal; and the gap between the two widens with every additional vendor relationship.

***P4:*** *The first-party accountability burden following a vendor incident increases as customer data is distributed across more third-party platforms. Each additional processor adds monitoring, assurance, coordination, and residual-risk costs that fall disproportionately on the focal organisation under incident-response time constraints.*

## 6. Discussion

The findings of this study explain why a third-party analytics incident can become a first-party governance problem. More specifically, our analysis shows that the OpenAI–Mixpanel incident is significant because it reveals how customer trust, vendor delegation, and data movement interact to create accountability beyond the technical boundary of a breach. This finding resonates with Chen and Jai (2019) study, which shows that data breaches can deplete customer trust by shaping perceptions of vulnerability, severity, and organisational response. It also extends Boyson et al. (2022) work on digital supply-chain risk by showing that third-party cybersecurity risk is also a trust and accountability problem. In AI service ecosystems, trust erosion may arise not only from the sensitivity of exposed data but also from the perceived legitimacy of delegated data flows. When customer data collected through a first-party AI service is transferred to a third-party analytics provider, accountability remains attached to the focal service provider because the customer relationship, data collection decision, and vendor delegation are all embedded in its governance arrangements. The findings, therefore, reflect a broader shift from incident-centred breach assessment to delegation-centred accountability, where the significance of a data incident depends not only on what data were exposed, but also on how responsibility is distributed across the vendor ecosystem.

The first finding that vendor relationships expand the effective attack surface should be interpreted as a boundary problem rooted in **transparency asymmetry**. In vendor-mediated digital supply chains, the focal organisation delegates data processing to external providers while remaining unable to fully observe the vendor's internal security posture, employee resilience to social engineering, privilege-separation practices, data-retention routines, or downstream dependencies. This creates a hidden-information risk in which the vendor controls operational conditions that shape exposure, yet these conditions remain only partially visible to the principal. As a result, the boundary of operational control does not necessarily coincide with the boundary of customer accountability. A vendor may control the system in which an incident occurs, yet the focal organisation may remain accountable to users because it selected the vendor, authorised the data transfer, and benefited from the delegated service. This interpretation is consistent with NIST's treatment of cybersecurity supply chain risk management as an enterprise governance concern, rather than a narrowly technical control function (National Institute of Standards and Technology, 2024). It is also aligned with broader C-SCRM guidance, which conceptualises supplier risk as a life-cycle issue spanning design, acquisition, operation, monitoring, and disposal (Baldwin, 2022). However, our findings extend this literature by showing that supply-chain risk is not confined to software dependencies, infrastructure compromise, or malicious code insertion. In analytics and metadata environments, even routine data sharing with vendors can create accountability risks. Once customer data moves outside the focal organisation's own systems, the organisation may not be able to fully verify how that data is protected, stored, or used. Thus, an organisation's cyber risk should not be understood only in terms of the systems it owns or operates. It should also include the vendor relationships through which customer data are shared and through which vendor-side security conditions become first-party accountability risks.
The second finding, concerning metadata, contributes to privacy and security theory by emphasizing adversarial actionability. The absence of credentials, API keys, payment details, or content clearly reduces incident severity. Nevertheless, account names, email addresses, browser information, location information, referring websites, and internal identifiers can support targeted phishing and organizational reconnaissance. This interpretation aligns with privacy scholarship showing that data sensitivity depends on context, linkage, and potential use rather than category alone (Belen-Saglam et al., 2022). It also explains why OpenAI's advisory warned users about phishing attempts even if passwords and API keys were not affected. A practical implication is that cybersecurity data classification should be expanded from a static hierarchy of data types to a threat-informed assessment of what an attacker can do with the data. Finally, the insufficiency of point-in-time assurance reflects the temporal nature of cyber risk. Agency theory predicts that delegation creates information asymmetry and monitoring costs (Eisenhardt, 1989). Vendor certifications, questionnaires, and onboarding reviews reduce this asymmetry, but they do not eliminate it. They are evidence of a control environment at a point in time, not proof of continuing control effectiveness. This finding is consistent with the NIST CSF 2.0 emphasis on monitoring and improving cyber supply chain risk management over the course of supplier relationships (National Institute of Standards and Technology, 2024) and with ISO/IEC 27001's orientation toward risk assessment, supplier relationships, and continual improvement (Kitsios et al., 2023). The important analytical point is not that every vendor requires constant surveillance, but that assurance intensity should track data sensitivity, service criticality, access privileges, and incident externality.

The final finding, data proliferation, identifies the mechanism through which ordinary business integrations create systemic exposure. Analytics tools, customer-support systems, and experimentation platforms are often introduced to improve product quality and operational learning. However, each additional data repository multiplies the number of places where identifiable metadata can be stored, exported, logged, retained, or misconfigured. This finding complements work on digital ecosystems and service specialization (Hofmann & Osterwalder, 2017) by showing that efficiency gains create governance costs when data flows are not tightly inventoried and minimized. It also explains why supply chain vulnerabilities have become a leading cyber-resilience concern among large organizations (World Economic Forum, 2025).

### 6.1 Reframing agency theory: from dyadic delegation to transitive cyber agency

The article’s first theoretical contribution is to extend agency theory from a dyadic model of delegation to a nested and transitive model of cyber governance. Classical agency theory explains how a principal governs an agent under information asymmetry through monitoring, bonding, and residual loss allocation (Jensen & Meckling, 1976). Third-party cybersecurity adds a structural complication: the end user who bears exposure is often not a party to the vendor contract, and the focal organization is simultaneously an agent of the user and a principal to vendors. This dual position explains why a vendor incident can become a first-party accountability event even when the focal organization's own systems are not compromised. More specifically, this study extends agency theory in three interrelated ways.

First, it shifts the unit of analysis from the isolated principal–agent dyad to a cascading agency chain, in which accountability and control are distributed across multiple actors: customers → focal organisation → vendors → vendor employees. This reconceptualisation is important because cybersecurity exposure is rarely produced by a single bilateral relationship; rather, it emerges through layered delegations in which downstream actors can materially affect upstream principals. Second, the study theorises the focal organisation's role duality. The focal organisation simultaneously acts as an agent accountable to customers and as a principal that governs vendors through contracts, audits, certifications, and evidence requests. This dual position creates a governance tension, as the organization is held accountable for outcomes generated within vendor environments that it can only partially observe and indirectly control. Recent empirical and regulatory developments substantiate this tension.

Cyberattacks propagate along dependency relations, imposing losses on organizations that extend well beyond those directly compromised (Crosignani et al., 2023), yet inter-firm cyber risk remains understudied relative to internal security measures (Ghadge et al., 2019). In response, regulators have likewise begun to codify this asymmetry, holding organizations fully accountable for functions outsourced to providers beyond their direct supervision, which has been summarised in Regulation (EU) 2022/2554 (Buneci, 2025). The governance tension identified here, therefore, can be seen as an emerging structural condition of digitally interdependent organizations. Third, the study introduces the concept of transitive residual loss, whereby losses arising from hidden information or actions of a downstream agent can be borne by an upstream organisation that did not directly control the compromised environment. These extensions are necessary for cybersecurity governance because customers rarely select, observe, or evaluate the gatekeepers that shape their security exposure; instead, accountability is mediated through the focal organisation that orchestrates delegated data flows.

The second theoretical contribution is to translate key concepts from agency theory into cybersecurity-governance mechanisms. P1 turns hidden information into a *visibility asymmetry* in which vendor attributes that cannot be fully verified become part of the effective attack surface. P2 turns an incomplete specification into an *actionability mismatch*, where contractual data classifications may miss how metadata can be weaponized. P3 turns hidden action into *assurance decay,* where a point-in-time signal loses value as vendor behaviour and operating conditions change. P4 turns agency costs into *accountability concentration*, in which monitoring, bonding, notification, remediation, and residual loss accumulate at the visible focal organization. In this way, the analysis specifies how agency problems materialize in cyber incidents.

### 6.2 The Fortress and Gatekeeper framework

Building on this extension of agency theory, the Fortress and Gatekeeper framework is the paper's integrative theoretical model. The fortress represents the visible organization that receives customer trust and bears customer-facing accountability. The gatekeepers represent vendors that hold data, privileges, infrastructure, analytics functions, or user-support roles. Figure 2 visualizes the framework as a *nested agency chain* where customer trust, data, and delegated functions move downstream from customers to the focal organization and then to gatekeepers and their own downstream agents; accountability, residual loss, notification burden, and trust repair move upstream toward the focal organization. The dashed boundary around the focal organization represents formal ownership, while the broader governance boundary follows the delegated trust and data pathways that determine exposure. The four propositions specify the framework's mechanisms: visibility asymmetry, actionability mismatch, assurance decay, and accountability concentration.

**Figure 2.** The Fortress and Gatekeeper framework: a nested agency model of transitive trust in third-party cybersecurity governance.

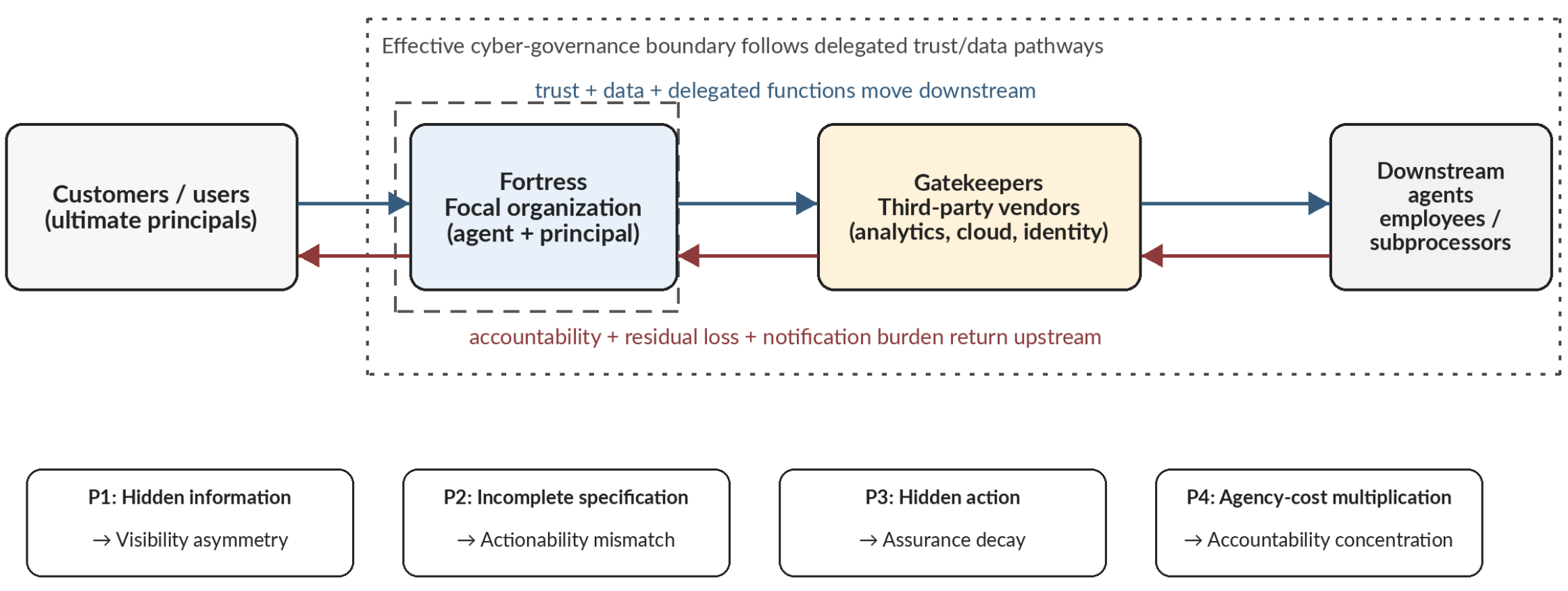

### 6.3 Boundary conditions

The refined framework should be interpreted within three boundary conditions. First, it is most applicable where end users primarily recognize and trust a single focal organization while the vendors supporting service delivery remain largely invisible. Its explanatory power may be weaker in co-branded, marketplace, or processor-visible settings where delegation is more transparent to users. Second, the present analysis is calibrated to incidents involving analytics and identifiable metadata. Incidents involving credentials, content, model assets, or safety-critical systems may trigger additional governance dynamics—such as more direct regulatory intervention or safety oversight—that lie beyond the current scope. Third, the framework assumes a largely hierarchical delegation chain characterized by asymmetric visibility between principal and agent. More reciprocal platform ecosystems or multilateral dependency structures may require extensions that draw on network-governance perspectives.

### 6.4 Governance implications

The framework also has practical implications. First, vendor tiering should be organized around trust transfer, data actionability, and customer-facing risk rather than contract value alone. Suppliers that process identifiable customer data, support critical user journeys, manage authentication, or hold privileged access should be treated as governance-critical vendors and reviewed accordingly. Second, data classification should become more threat-informed. The case shows that metadata can enable phishing, reconnaissance, impersonation, and linkage even when it falls outside conventional categories of highly sensitive information. Organizations should therefore classify vendor-held data not only by legal sensitivity, but also by adversarial actionability. Third, contracts should be treated as operational security instruments rather than as purely commercial documents. Provisions on notification timelines, evidence sharing, log preservation, audit rights, subcontractor controls, deletion obligations, and termination triggers shape what the focal organization can actually do under incident conditions. Fourth, high-risk vendors require ongoing rather than episodic assurance. Continuous or event-driven reviews, access and authentication checks, tabletop exercises, and clear escalation triggers are necessary to reduce assurance decay over the life of the relationship. These implications also reinforce the importance of data minimization and pathway reduction: the fewer external copies, processors, and routes that exist, the lower the accountability burden when a vendor incident occurs. Table 3 translates the updated framework into operational terms.

Table 3. Agency-theory extension, framework dimensions, and governance implications

| Agency-theory construct | Framework dimension | Governance implication |
|---|---|---|
| Hidden information / adverse selection | P1, Visibility asymmetry | Map vendors by customer-facing services, data access, privilege level, and unverifiable control attributes. |
| Incomplete specification / residual loss | P2, Actionability mismatch | Classify data by adversarial actionability, not only by legal or contractual sensitivity labels. |
| Hidden action / moral hazard | P3, Assurance decay | Use continuous or event-driven assurance, evidence refreshes, notification tests, access reviews, and incident simulations. |
| Monitoring, bonding, and residual-loss multiplication | P4, Accountability concentration | Reduce external data copies and pathways; maintain current data-flow inventories, playbooks, and vendor exit options. |

### 6.5 Limitations and future research

Several limitations should temper interpretation. First, this is a single-case document analysis, so the goal is analytical generalization rather than prevalence estimation. Second, the study relies on public disclosures and published standards, which are necessarily partial and may reflect legal, reputational, and strategic constraints. Third, the analysis cannot observe internal contracts, control evidence, forensic artifacts, or decision processes beyond what has been publicly reported. Fourth, because the case centres on analytics-related metadata rather than content, credentials, or model assets, the framework has not yet been tested against incidents involving more sensitive or technically consequential forms of exposure. These limitations point to a clear research agenda. Comparative case studies could examine whether the *Fortress and Gatekeeper framework* travels across analytics, cloud, identity, managed-service, and AI supply-chain incidents, and whether the four mechanisms identified here, including *visibility asymmetry, actionability mismatch, assurance decay, and accountability concentration*, appear in different governance contexts. Interview-based research with CISOs, procurement leaders, privacy officers, legal counsel, and product teams could clarify how organizations actually perform vendor tiering, data classification, scope determination, customer notification, and trust repair after supplier incidents. Quantitative work could operationalize transitive-trust exposure—for example, through the number of vendors holding identifiable customer data, the concentration of critical workflows in third parties, the actionability of vendor-held data, or the time required to identify affected users after a vendor incident—and test how these measures relate to notification speed, remediation cost, or downstream customer loss over time.